\documentclass[aps,prd,unsortedaddress,superscriptaddress,showpacs,nofootinbib]{revtex4-2}
\usepackage{graphicx,epsfig,epstopdf}
\usepackage{amssymb,amsmath,amsxtra,amsfonts,latexsym}
\usepackage{cancel}
\usepackage{bm}
\usepackage{float}
\usepackage{longtable}
\usepackage{multirow}
\usepackage{booktabs}
\usepackage{array}
\usepackage{wrapfig}
\usepackage{color}
\usepackage{slashed}
\usepackage{dsfont}
\usepackage{bbm}
\usepackage{titlesec}
\usepackage [latin1]{inputenc}

\titleformat{\section}{\large\bfseries}{\thesection}{1em}{}

\newcommand{\bea}{\begin{eqnarray}}
\newcommand{\ena}{\end{eqnarray}}
\newcommand{\be}{\begin{equation}}
\newcommand{\en}{\end{equation}}
\newcommand{\nn}{\nonumber\\}
\newcommand{\Tr}{\mbox{\rm{tr}}}

\newcommand{\Jpsi}{\ensuremath{J\!/\!\psi}}

\begin{document}

\title{Hidden-charm strong decays of the spin-$2$ partner of $X(3872)$}

\author{Gurjav Ganbold}
\email{ganbold@theor.jinr.ru}
\affiliation{Bogoliubov Laboratory of Theoretical Physics,
Joint Institute for Nuclear Research, 141980 Dubna, Russia}
\affiliation{Institute of Physics and Technology, Mongolian Academy
of Sciences, 13330 Ulaanbaatar, Mongolia}

\author{M. A. Ivanov}
\email{ivanovm@theor.jinr.ru}
\affiliation{Bogoliubov Laboratory of Theoretical Physics,
Joint Institute for Nuclear Research, 141980 Dubna, Russia}

\begin{abstract}
Hidden-charm strong decays of the spin-2 partner $X_2(4014)$ of the
charmonium-like state $X(3872)$ have been studied in the framework of the
covariant confined quark model. The exotic state $X_2$ has been interpreted
as a four-quark state with molecular-type interpolating current. We have
considered the decay widths of $X_2$ on the level of two-petal quark loops.
The partial widths of the strong decays $X_2\!\to\!\omega \, J/\Psi$ and
$X_2\!\to\!\rho^0 \, J/\Psi$ have been calculated and the related branching
ratio has been analyzed.



\end{abstract}

\maketitle

\section{Introduction}
\label{sec:intro}

During the last two decades, a large number of new mesonic structures was
discovered in the course of experimentally establishing the heavy hadron
spectrum. The positions of many of them in the mass spectrum, quantum
numbers, and decay widths seem to not fit the predictions of the conventional
quark models of mesons (quark-antiquark pairs) and baryons (three-quark
structures). These 'exotic' states are commonly referred to as $XYZ$  states.

The massive amount of data accumulated over the past years on $XYZ$ states
makes the theoretical study of exotic hadrons with 'charm' quarks an essential
tool for analyzing quantum chromodynamics at the confinement scale.
Nowadays, a debate is ongoing between theoretical models about the nature of
$XYZ$ hadrons, and assumptions about different multiquark configurations have
been developed to explain their  properties, such as mass, decay width, and
the main quantum numbers (for reviews see, e.g., Refs.~\cite{Chen:2016qju,
Esposito:2016noz,Guo:2017jvc,Brambilla:2019esw,Chen:2022asf}).

The first member of the $XYZ$ family, the $X(3872)$  state in the charm sector
was observed by the Belle collaboration in 2003 as a structure in the
$\pi^+\pi^-J/\psi$ invariant mass spectrum \cite{Belle:2003nnu}. Later, the
LHCb Collaboration in 2013 determined its quantum numbers $J^{PC}=1^{++}$
\cite{LHCb:2013kgk}. The $X(3872)$ is located very close to the $D\bar{D}^*$
threshold, and to date, $X(3872)$ is naturally considered as a shallow bound
mesonic molecule  $(D\bar{D}^*)$ with quantum numbers $J^{PC} = 1^{++}$
(see, e.g.  Refs.~\cite{flem:2008prd,mehen:2015prd,Kalashnikova:2018vkv,
meng:2021prd,wang:2022prd}).

Furthermore, the $X(3872)$ is often used a basis for predicting possible exotic
states due to the achievement of its molecular picture.

A possible existence of the heavier partner of $X(3872)$ with a similar value
for the binding energy and mass $M=4015$~MeV was first predicted in
~\cite{Tornqvist:1993ng}. Then, such a state with $D^*\bar{D}^*$ molecular
structure has been guessed as a consequence of the heavy-quark spin symmetry
for the system under consideration in~\cite{Nieves:2012tt}. Later,  the existence
of an isoscalar $2^{++}$ $D^*\bar{D}^*$ partner of the  $X(3872)$ has been
predicted in a number of investigations using various phenomenological models
(e.g., in Refs.~\cite{Molina:2009ct, Nieves:2012tt, hidalgo:2013plb,
albaladejo:2015epjc, baru:2016plb, Ortega:2017qmg, Dong:2021juy, shi:2023}).

Recently, from the experimental side, the Belle collaboration has reported a
hint of an isoscalar structure with mass $M=(4014.3 \pm 4.0 \pm 1.5)$ MeV and
width $\Gamma=(4\pm 11 \pm 6)$ MeV, seen in the $\gamma \psi(2S)$ invariant
mass distribution via a two-photon process \cite{Belle:2021nuv}.

This new structure is located near the $D^*\bar{D}^*$ threshold, so one may
conclude that it is a promising candidate for the corresponding loosely bound
state. In particular, this structure was assumed to be a $D^*\bar D^*$ molecule
with $J^{PC}=0^{++}$ in~\cite{Yue:2022gym,Duan:2022upr}. Furthermore, its
measured width has the same order of magnitude as the prediction in
~\cite{Albaladejo:2015dsa, Baru:2016iwj}. Nowadays, this narrow state is indeed
a potential candidate for a $(D^* \bar{D}^*)$ molecule with $J^{PC}=2^{++}$.

Alternatively, within the conventional pattern of mesons, a $2^{++}$ tensor
state with a similar mass could also be a conventional charmonium state in
the first radial excitation $\chi_{c2}(3930)$ (see, e.g.
Refs.~\cite{Godfrey:1985xj,Li:2009ad}).

Also, a compact tetraquark model has been employed to explore the $2^{++}$
states in~\cite{Wu:2018xdi,Shi:2021jyr,Giron:2021sla}.

In such a situation, the seemingly easiest way to disentangle these different
multiquark configurations may be just to analyze the mass splitting between the
$2^{++}$ and $1^{++}$ states. According to~\cite{Nieves:2012tt,Guo:2013sya},
the corresponding mass splitting is approximately equal to that between the
vector and pseudoscalar charmed mesons, that leads to a difference
$\triangle M = M_{X_2(4014)} - M_{X(3872)} \approx 142~\textrm{MeV}$.
However, the tetraquark approach  predicted a much smaller  difference
$\triangle M  \approx 80~\textrm{MeV}$~\cite{Maiani:2014aja}.
For the difference between the charmonium $\chi_{c2}(3930)$ with $2^{++}$
and $X(3872)$  a similar conclusion has been found within the Godfrey-Isgur
quark model~\cite{Godfrey:1985xj} and a screened potential
model~\cite{Li:2009ad} - but the corresponding differences $\triangle M$ are
even less - about only $30$~MeV and $40$~MeV, respectively.

Another more complicated, but more reliable way to discriminate the various
multiquark configurations of the spin-$2$ partner of the $X(3872)$ is to
investigate the decay properties of the $X_2(4014)$ tensor state.
In particular, a quark model in~\cite{Barnes:2005pb} provides estimates for the
$X_2$ decay width to charmed mesons around tens of MeV by considering a
$2^{++}$ tensor structure as the first radial excitation of the $P$-wave
$\chi_{c2}(2^3P_2)$ charmonium. Then, the hadronic decays of the  $S$-wave
$D^*\bar{D}^*$ hadronic molecule, into $D\bar{D}$ and $D\bar{D}^*$ meson pairs
were estimated to be small of the order of a few MeV ~\cite{Albaladejo:2015dsa}
and, vise versa, as large as  50~MeV~\cite{Baru:2016iwj}.

In a series of our previous papers we have studied some of the $XYZ$-states in
the framework of the covariant confined quark model (CCQM) (for review, see
e.g., \cite{Ivanov:2018ayq,Dubnicka:2020yxy,Ganbold:2024epja}).  The CCQM
implements effective quark confinement realized by introducing an infrared
cutoff of the proper time  integration that prevents any singularities in matrix
elements of the hadronic processes. In a number of papers, we have applied
the CCQM to calculate the various physical observables like the form factors,
angular decay characteristics, decay branchings etc. in the processes involving
both light and heavy mesons and baryons.
In particular, inspired by recent measurements we have studied the radiative
decays of charmonium  states below the $D {\bar D}$ threshold by introducing
only one adjustable parameter common for the six charmonium
states~\cite{Ganbold:2021nvj}. The obtained results were in good agreement
with the latest data. We also predicted a narrower full width for the $h_c$
charmonium than reported in PDG-2020.

We have treated the first exotic $X(3872)$-meson as a diquark-antidiquark bound
state and calculated its strong and electromagnetic decays
in~\cite{Dubnicka:2010kz,Dubnicka:2011mm}. The four-quark structure of the
charged $Z_c(3900)$, $Z(4430)$, $X_b(5568)$, $Z_b(10610)$ and $Z_b'(10650)$
states has been examined by us in ~\cite{Goerke:2016hxf,Goerke:2017svb}.
The widths of the strong two-body decays $Z_c^+ \to J/\psi\pi^+$,
($\bar D^{(\ast)}D$) and  many others have also been calculated. It was found
that the tetraquark-type current widely used in the literature for the
$Z_c(3900)$ leads to a significant suppression of the $\bar D D^\ast$ and
$\bar D^\ast D$ modes. Contrary to this a molecular-type current provides an
enhancement by a factor of 6-7 for the $\bar D D^\ast$ modes compared with
the $Z_c^+\to J/\psi\pi^+$, $\eta_c\rho^+$ modes, which is in agreement with
experimental data from the BESIII Collaboration.

The $Y(4230)$ resonance has been analyzed in~\cite{Dubnicka:2020xoh} as a
four-quark state. Two options for the interpolating currents have been studied
by using either the molecular-type current or tetraquark one. In both cases the
widths of two-body decays $Y(4260)\to Z_c(3900)+\pi$ and
$Y(4260)\to D^{(\ast)}+\bar D^{(\ast)}$ have been calculated. It was found that
in both approches the mode $Y\to Z^+_c + \pi^-$ is enhanced compared with
the open charm modes.
More recently, strong decays of the charmonium-like state $Y(4320)$ have
been studied within the CCQM. The resonance $Y$ has been interpreted  as a
four-quark state  with molecular-type interpolating current. We  evaluated the
hidden-charm decay width of $Y$ into a vector and a scalar, with the latter
decaying subsequently to a pair of charged pseudoscalar states. The strong
decay mode $Y \!\to\! \pi^{+}\pi^{-}J/\Psi$ has been studied by involving the
scalar resonances $f_0(500)$ and $f_0(980)$, considered quark-antiquark
states, while the mode $Y \!\to\! K^{+}K^{-} J/\Psi$  - via $f_0(980)$. We have
calculated the partial widths of the related strong decays and the branching
ratio ${\cal B}(Y \!\to\! K^{+}K^{-} J/\Psi)$/${\cal B}(Y \!\to\! \pi^{+}\pi^{-}J/\Psi)$,
recently determined by the BESIII collaboration. The estimated branching ratio
and calculated partial strong decay widths were in reasonable agreement with
thelatest experimental data \cite{Ganbold:2024epja}.

In the light of these our findings, we below consider the $X_2(4014)$ state as
a four-quark state of the molecular-type. We investigate the strong two-body
decays $X_2 \!\to\! \omega J/\Psi$ and $X_2 \!\to\! \rho^0 J/\Psi$ in the
framework of the CCQM.

The paper is organized as follows.
A brief introduction to the CCQM and the general formalism for describing
$X_2(4014)$ as four-quark state with molecular-type current are given in
Section~\ref{sec:approach}. In Section~\ref{sec:strong} we determine the
amplitudes and partial widths of the two-body strong decays
$X_2 \!\to\! \omega J/\Psi$ and $X_2 \!\to\! \rho^0 J/\Psi$.  We discuss the
obtained results of calculation and compare them with latest theoretical
predictions in Section~\ref{sec:numerical} . Our findings are summarized in
Section~\ref{sec:summary}.

\section{Approach}
\label{sec:approach}

The CCQM~\cite{Branz:2009cd} represents an effective quantum field approach
to hadronic physics and it is based on a relativistic Lagrangian describing the
interaction of a hadron with its constituent quarks. It is a universal, pure relativistic,
and a manifestly Lorentz covariant approach and allows one to study bound states
with an arbitrary number of constituents and with arbitrary quantum numbers
(spin-parity, isospin, flavor content, etc.)
~\cite{bran10ivan17,Gutsche:2018nks, Dubnicka:2015iwg,
Gutsche:2015mxa,Gutsche:2013pp,Ivanov:2015woa}.
Therefore, this model may serve an appropriate theoretical framework to analyze
the strong decays of the exotic $X_2(4014)$ state.

According to the CCQM, a hadron described by a field $H (x)$ is coupled to
a non-local quark current $J_H$ carrying the hadron quantum numbers by the
interaction Lagrangian.

In particular, the effective interaction Lagrangian, describing the coupling
of the exotic state $H=X_2(4014)$,  to its constituent four quarks may be
written in the form:
\be
{\cal L}_{\rm int} = g_{H}\, H_{\mu\nu}(x)\cdot J^{\mu\nu}_{H}(x) + \text{H.c.}\,,
\label{eq:lagran}
\en
where $g_{H}$ is the renormalization coupling of the hadron $H$.

The interpolating four-quark molecular-type current for the neutral state
$X_2(4014)$ with the quantum numbers $I^G(J^{PC}) = 0^{+}(2^{++})$ may be
introduced as follows:
\be
J_H^{\mu\nu} = \frac{1}{\sqrt{2}}
\left\{ (\bar{q} \gamma^\mu c) (\bar{c} \gamma^\nu q)
      + (\gamma^\mu \leftrightarrow \gamma^\nu) \right\}
\label{eq:X2cur}.
\en
The corresponding nonlocal generalization of the four-quark current within the
CCQM reads
\bea
\label{eq:X2nonloc}
J_H^{\mu\nu} (x) &=& \int\! dx_1\ldots \int\! dx_4 \delta
\left( x-\sum\limits_{i=1}^4 w_i x_i \right)
\Phi_{\,H}\Big(\sum\limits_{i<j} (x_i-x_j)^2 \Big)
J_{H_{non}}^{\mu\nu} (x_1,\ldots,x_4),   \\
J_{H_{non}}^{\mu\nu} &=&  \tfrac{1}{\sqrt{2}} \Big\{
 (\bar q(x_3) \gamma^\mu c(x_1))\cdot (\bar c(x_2) \gamma^\nu q(x_4) )
+ (\gamma^\mu \leftrightarrow \gamma^\nu)\Big\},
\qquad (q=u,d),   \nonumber
\ena
where the reduced quark masses $w_i=m_i/ \left( \sum_{j=1}^4 m_j \right)$ are
specified for $m_1=m_2 = m_c$ and $m_3=m_4 = m_q$. Hereby, we neglect the
isospin violation in the $u-d$ sector, i.e. $m_u=m_d=m_q$. The numbering of
the coordinates $x_i$ is chosen such that one has a convenient arrangement of
vertices and propagators in the Feynman diagrams to be calculated.

The translationally invariant four-quark non-local vertex function $\Phi_H$ in
Eq.~(\ref{eq:X2nonloc}) characterizes the quark distribution inside the hadron
and reads:
\be
\Phi_H \Big(\sum\limits_{i<j} (x_i-x_j)^2 \Big) =
\prod\limits_{i=1}^3\int\!\!\frac{d^4 q_i}{(2\pi)^4}\, \widetilde\Phi_H (- Q^2) \,
e^{ - i q_i ( x_i - x_4 ) } \,,
  \qquad  Q^2 \doteq \frac12\sum\limits_{i\le j}q_iq_j \,.
\label{eq:X2vertex}
\en
The Fourier transform of the translational invariant vertex function in momentum
space is required to fall off in the Euclidean region in order to provide the
ultraviolet convergence of the loop integrals. The vertex function
$\widetilde{\Phi}_H\left(-Q^2\right)$ is unique for the given hadron $H$ and
each hadron has its own adjustable parameter $\Lambda_H$, which can be
related to the hadron 'size'. For most cases a pattern can be traced - the
heavier a hadron, the larger its 'size'.

Below we use a simple Gaussian form as follows:
\be
\widetilde\Phi_{H}(- Q^2) = \exp(Q^2 / \Lambda^2_{H} ) \,.
\label{eq:X2vertsize}
\en
In fact, any choice for  $\widetilde\Phi_{H}$ is appropriate as long as
it falls off sufficiently fast in the ultraviolet region to render the corresponding
Feynman diagrams ultraviolet finite.

According to the CCQM, the renormalization coupling $g_{H}$ of a hadron
in Eq.~(\ref{eq:lagran}) should be determined according to the so-called
'compositeness condition'~\cite{Salam:1962ap,Weinberg:1962hj}
which imposes that the renormalization constant of the hadron wave function
has to be equal to zero as follows:
\be
Z_H = 1 -  g_H^2 \frac{d}{dp^2} \, \tilde{\Pi}_H(p^2)    = 0 \,,
\qquad p^2=M_H^2 \,,
\label{eq:renorm}
\en
where $M_H$ is the hadron mass and  $\tilde{\Pi}_H( p^2 )$ is the diagonal
(scalar) part of the hadron self-energy.
The requirement $Z_H=0$ implies that the physical state does not contain
the bare state and is appropriately described as a bound state. The interaction
leads to a dressed physical particle, i.e. its mass and wave function have to be
renormalized. The condition $Z_H=0$ also excludes effectively the constituent
degrees of freedom from the space of physical states. It thereby guarantees
the absence of double counting for the physical observable under consideration,
the constituents only exist in virtual states.

\begin{figure}[h]
\vspace{-5mm}
\centering
\includegraphics[width=0.6\textwidth]{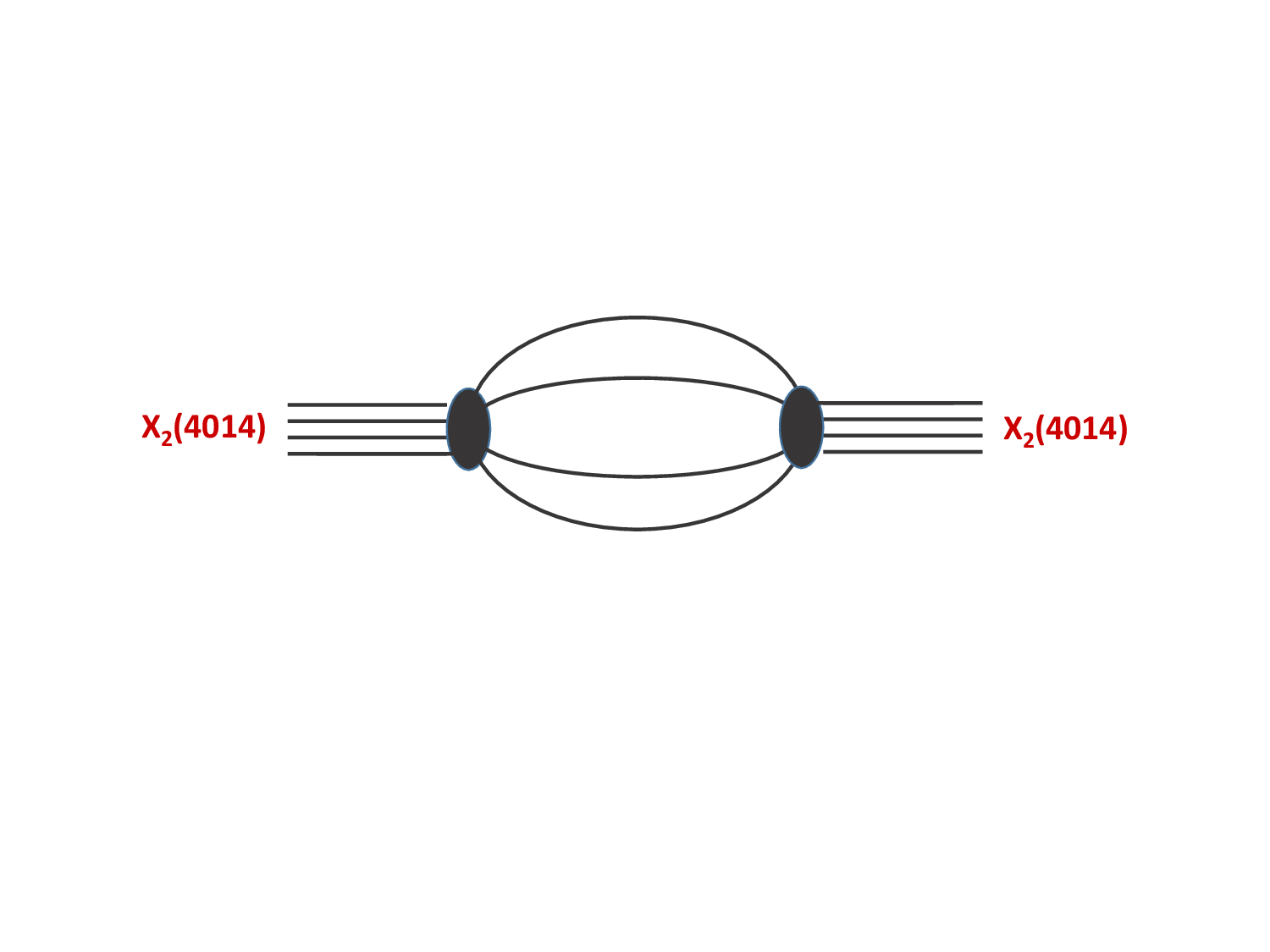}
\vspace{-15mm}
\caption{Feynman diagram for the $X_2(4014)$ mass operator.}
\label{fig:FIG1}
\end{figure}

In particular,  the mass operator for the four-quark molecularly structured
exotic state $X_2(4014)$ depicted in Fig.~\ref{fig:FIG1} reads
\bea
\widetilde\Pi^{\mu\nu\rho\sigma}_{X_2}(p)
&=&
\frac{N_c^2}{2} \,\prod\limits_{i=1}^3\int\!\!\frac{d^4k_i}{(2\pi)^4i}\,
\widetilde\Phi_{X_2}^2
\left( -(k_1^2 + k_2^2 + k_3^2 - k_1 k_2  - k_1 k_3 + k_2 k_3)/2 \right)
\nonumber\\
&\times&
\Big\{
\Tr\left[ S_3(\not \!{k}_3 + w_3 \not \!{p})\gamma^\mu
             S_1(\not \!{k}_1-  w_1 \not \!{p}) \gamma^\rho  \right]
\nonumber\\
&&
\times \Tr\left[ S_2(\not \!{k}_2 + w_2 \not \!{p})\gamma^\nu
             S_4(- \not \!{k}_1 + \not \!{k}_2 + \not \!{k}_3 - w_4 \not \!{p})
             \gamma^\sigma \right]
\nonumber\\
&&
+ \,
\Tr\left[ S_3(\not \!{k}_3 + w_3 \not \!{p})\gamma^\mu
             S_1(\not \!{k}_1-  w_1 \not \!{p}) \gamma^\sigma  \right]
\nonumber\\
&&
\times \Tr\left[ S_2(\not \!{k}_2 + w_2 \not \!{p})\gamma^\nu
             S_4(- \not \!{k}_1 + \not \!{k}_2 + \not \!{k}_3 - w_4 \not \!{p})
             \gamma^\rho  \right] \Big\} \,,     \quad   N_c = 3 \,.
\label{eq:self}
\ena

The corresponding diagonal (scalar) part is defined as follows:
\be
\widetilde\Pi_{X_2}(p^2) = \frac{1}{5}
\left(
 \frac{1}{2} \bar{g}_{\mu\rho} \bar{g}_{\nu\sigma}
+ \frac{1}{2} \bar{g}_{\mu\sigma} \bar{g}_{\nu\rho}
- \frac{1}{3} \bar{g}_{\mu\nu} \bar{g}_{\rho\sigma}
\right)
\widetilde\Pi^{\mu\nu\rho\sigma}_{X_2}(p)\,,
\quad
\bar{g}_{\mu\nu} \doteq  - g_{\mu\nu} + \frac{p_\mu p_\nu}{p^2} \,.
\label{eq:self0}
\en

For the quark propagator we use the Fock-Schwinger representation:
\be
\widetilde{S}_j(\not \!{k}) =  \left( m_j + \not \!{k} \right)
\intop_{0}^{\infty}
d\alpha\: \exp\big(-\alpha \left(m_j^2 - k^2 \right) \big) \,.
\label{eq:Fock}
\en

Further on, we consider the hadron mass operator, matrix elements of hadronic
decays and transitions, which are represented by quark-loop diagrams, which are
described as convolutions of the corresponding quark propagators and vertex
functions.

Then, the integration over quark-loop momenta can be performed for arbitrary
Feynman diagram with n-quark propagators. The final result will be the
multidimensional integration over the Fock-Schwinger parameters. Such integral
contains all information on the analytical structure of the Feynman diagram
including the branch points corresponding to the threshold singularities. These
singularities lead to the imaginary part of the diagram if the ingoing energy
larger than sum of quark masses.
Physically this means that the quarks may appear in the observed spectrum.

In order to solve  the quark confinement problem the following trick has been
used. First, one has to  transform the integral over an infinite space of the
Fock-Schwinger parameters into an integral over a simplex convoluted with only
one-dimensional improper integral. One has
\bea
\Pi
&=&
\intop_0^\infty\!\! d^{\,n}\alpha\: W(\alpha_1,\ldots,\alpha_n)
= \intop_0^\infty\!\! d^{\,n}\alpha\: \underbrace{\intop_0^\infty\! dt\: \delta
\left(t-\sum_{i=1}^{n}\alpha_{i}\right)}_{={1}} W(\alpha_1,\ldots,\alpha_n)
\nn
&=&
\intop_0^\infty\! dt\:  t^{n-1}\intop_0^\infty\!\! d^{\,n}\alpha\:
\delta\left(1-\sum_{i=1}^{n}\alpha_{i}\right) W(t\alpha_1,\ldots,t\alpha_n),
\label{eq:trick}
\ena
where the integration over dimensionless $\alpha_i$ parameters proceeds
over a simplex. Now there is the only dimensional parameter ''$t$'' which is
nothing as the Fock proper time.

The cut-off parameter $\lambda$ is then introduced in a natural way as follows:
\be
\intop_{0}^{\infty}dt\: t^{n-1}\ldots\rightarrow
\intop_{0}^{1/\lambda^{2}}dt\:t^{n-1}\ldots.
\label{eq:cut}
\en
Such a cut-off makes the integral to be an analytic function without any
singularities. In this way all potential thresholds in the quark-loop diagrams
are removed together with corresponding branch points \cite{Branz:2009cd}.
Within the CCQM, the cut-off parameter is universal for all processes and its
value, as obtained from a fit to data, equals to
\be
\mathrm{\lambda_{cut-off}=0.181\,\text{GeV}.}
\label{eq:cut-value}
\en
The resulting integrals are computed numerically.

The CCQM consists of several basic parameters: the universal infrared cutoff
parameter $\lambda$, the constituent quark masses $m_q$ and the hadron 'size'
parameters $\Lambda_H$. For given value of the size parameter $\Lambda_H$
the coupling constant $g_H$ of hadron $H$ is strictly fixed by the requirements
of Eq.~(\ref{eq:renorm}) and do not constitute further free parameters.
The model parameters are determined by minimizing $\chi^2$ in a fit to the
latest available experimental data and some lattice results. In doing so, we
have observed that the errors of the fitted parameters are of the order of
$\sim 10 \%$.
The central values of the basic parameters updated  in Refs.
~\cite{Ganbold:2014pua,Gutsche:2015mxa} are shown in Table~\ref{tab:TAB3}.
Obviously, the errors of our calculations within the CCQM are expected to be
about $\pm 10 \%$.

Below we apply the CCQM to estimate the strong decays of the spin-2 partner
$X_2(4014)$ of the charmonium-like state $X(3872)$.

\section{Strong decays of $X_2(4014)$ into $\omega \Jpsi$ and $\rho^0 \Jpsi$}
\label{sec:strong}

The ratio of the branching fractions
\be
\mathrm{BR_{X_2}}   \doteq
\frac{\Gamma(X_2 \to \omega J/\psi)} {\Gamma(X_2 \to \rho^0 J/\psi)}
\label{eq:BRstr}
\en
has recently been investigated using the effective Lagrangian approach  by assuming the $X_2$ as a molecular state of $D^*\bar{D}^*$ \cite{zheng:2024prd}.
The only contributions from the triangle hadron loops made of the charmed
mesons  $D^*$ and $\bar{D}^*$ have been considered.

It has been found that the decay widths are quite sensitive to the $X_2$ mass.
At the present center mass $M_{X_2}=4.0143~\mathrm{GeV}$, the width for
the $X_2\to J/\psi \rho^0$ was a few tens of keV, while it is on the order
of $10^{2}-10^{3}$ keV for the $X_2\to J/\psi \omega$.
The corresponding width ratio was calculated to be
\be
\mathrm{BR_{X_2}}  \approx 15 \,,
\label{eq:rat15}
\en
i.e. one order of magnitude larger than that for the case of $X(3872)$ which
approaches unity.

\begin{figure}[h]
\centering
\includegraphics[width=0.6\textwidth]{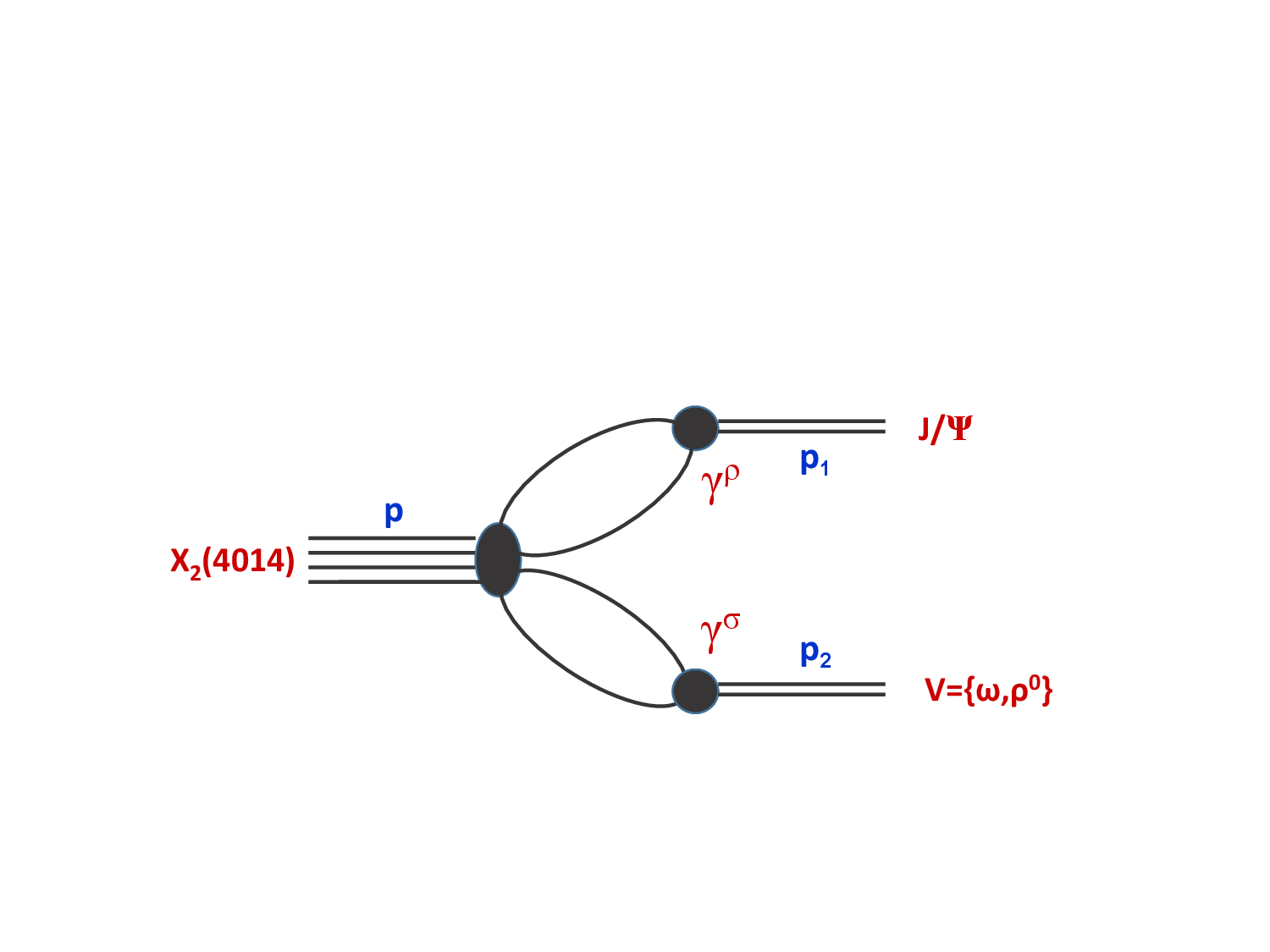}
\caption{Feynman diagram describing the decay $X \to \Jpsi + V$,
here $V=\{\omega,\rho^0\}$. }
\label{fig:FIG2}
\end{figure}

Below, we consider the $X_2(4014)$ as a four-quark state with a molecular-type
interpolating current and study its strong decays into $\omega \, J/\psi$ and
$\rho^0 \, J/\psi$ in the framework of the  CCQM~\cite{Branz:2009cd}. In doing
so, we limit ourselves by considering only the LO contributions corresponding
to the two-petal quark loops represented by the Feynman diagram in
Fig.~\ref{fig:FIG2}. We calculate the partial widths of  the related strong
decays and estimate the branching ratio mentioned in Eq.~(\ref{eq:BRstr}).

We note that the description of the $c\bar{c} - q\bar{q}$ transitions, which go
via gluon exchange, is out of the CCQM scope.

The invariant matrix element for the strong decay $X_2 \to \Jpsi + V$ reads
\bea
{\cal M}_{X_2 JV} \equiv  i \, (2\pi)^4\delta^{(4)}(p-p_1-p_2) \,
\varepsilon_{\mu\nu}(p)  \, \varepsilon^{\ast}_{\rho}(p_1) \,
\varepsilon^{\ast}_{\sigma}(p_2) \, \, T_{X_2 JV}^{\mu\nu\rho\sigma} (p_1,p_2) \,,
\label{eq:matrix}
\ena
where $\{p, p_1, p_2\}$ and $\varepsilon_{\mu\nu}(p)$,
$\varepsilon^{\ast}_{\rho}(p_1)$, $\varepsilon^{\ast}_{\sigma}(p_2)$ are the
momenta and polarization vectors of the $X_{2}$, $\Jpsi$ and the vector
meson $V=\{\omega,\rho^0\}$, correspondingly.

The polarization vectors of the tensor and vector mesons satisfy the symmetry,
transversality, tracelessness, orthonormality and completeness conditions as
follows:
\bea
&&
\varepsilon^{(\lambda)}_{\mu\nu}(p) = \varepsilon^{(\lambda)}_{\nu\mu}(p),
\qquad
\varepsilon^{(\lambda)}_{\mu\nu}(p)\, p^\mu = 0,
\qquad
\varepsilon^{(\lambda)}_{\mu\nu}(p)\, g^{\mu\nu} = 0,
\qquad
\varepsilon^{\dagger\,(\lambda)}_{\mu\nu} \varepsilon^{(\lambda')\,\mu\nu}
=\delta_{\lambda \lambda'} \,,
\nonumber\\
&&
\widetilde{g}_{\mu\nu\alpha\beta} \doteq
\sum\limits_{\lambda=0,\pm 1, \pm 2} \!\!\!
\varepsilon^{(\lambda)}_{\mu\nu}
\varepsilon^{\dagger\,(\lambda)}_{\alpha\beta}
=  
\frac{1}{2} \left(\bar{g}_{\mu\alpha}\,\bar{g}_{\nu\beta}
         + \bar{g}_{\mu\beta}\,\bar{g}_{\nu\alpha}\right)
- \frac{1}{3}\,\bar{g}_{\mu\nu}\,\bar{g}_{\alpha\beta} \,,
\quad
\bar{g}_{\mu\nu} \doteq - \, g_{\mu\nu}+\frac{p_\mu\,p_\nu}{p^2} \,,
\nonumber\\
&&
\epsilon^{(\lambda)}_\mu(p) \, p^\mu = 0 \,,
\quad
\sum\limits_{\lambda=0,\pm}\epsilon^{(\lambda)}_\mu(p)
\epsilon^{\dagger\,(\lambda)}_\nu (p) =\bar{g}_{\mu\nu} \,,
\quad
\epsilon^{\dagger\,(\lambda)}_\mu \epsilon^{(\lambda')\,\mu} 
= - \delta_{\lambda \lambda'} \,.
\label{polarX2}
\ena

In the leading order, the decay amplitude in Eq.~(\ref{eq:matrix}) reads
\bea
&&
\!\!\!\! \!\!\!\! T_{X_2 JV}^{\mu\nu\rho\sigma} (p_1,p_2)
= g_{X_2} \, g_{\Jpsi} \, g_{V} \,  \nn
&\times&
 \frac{N_c^2}{2}
\int\!\!\frac{d^4k_1}{(2\pi)^4i}\,\int\!\!\frac{d^4k_2}{(2\pi)^4i}\,
\widetilde\Phi_{X_2}\left(-Q^2\right)
\widetilde\Phi_{\Jpsi}\left(-\,(\ell_1 + \ell_2)^2/4\right)
\widetilde\Phi_{V}\left(-\,(\ell_3 + \ell_4)^2/4\right) \nn
&\times&
\Big\{
\Tr\left[\gamma^\mu S_1(\ell_1) \gamma^\rho S_2(\ell_2)
         \gamma^\nu S_4(\ell_4) \gamma^\sigma S_3(\ell_3) \right]
+
\Tr\left[\gamma^\nu S_1(\ell_1) \gamma^\rho S_2(\ell_2)
         \gamma^\mu S_4(\ell_4) \gamma^\sigma S_3(\ell_3) \right]
\Big\} \,,
\label{eq:amplitude}
\ena
with the following notations introduced:
\bea
\widetilde\Phi_H(- k^2) &=& \exp(k^2/\Lambda^2_H), \qquad H=\{X_2,\Jpsi,V\} \,,
\nn
Q^2 &=& \left[ (\ell_1+p w_1)^2+ (\ell_2-p w_2)^2+(\ell_3+p w_3)^2
  +(\ell_4-p w_4)^2\right]/2,
\nn
\ell_1 &=& k_1 - p w_1, \quad \ell_2 = k_1 - p w_1 - p_1,
\quad \ell_3 = k_2 - p w_4 - p_2, \quad \ell_4 = k_2 - p w_4 \,.
\nonumber
\ena

As mentioned above, the renormalized couplings $g_{X_2}$,  $g_{\Jpsi}$ and
$g_{V}$ are strictly determined  by the self-energy (mass) function of the
corresponding hadrons, (see Eq.~(\ref{eq:renorm})).

By substituting the corresponding Gaussian-type vertices functions and the
quark propagators in Eq.~(\ref{eq:amplitude}), and performing an explicit
$k_{1,2}$-integrations while turning the set of Fock-Schwinger parameters
into a simplex, we obtain the LO amplitude of the strong decay
$X_2 \to \Jpsi + V$ as follows:
\bea
T_{X_2 JV}^{\mu\nu\rho\sigma}(p_1,p_2)
&=&
A_V\cdot \Big( g^{\mu\rho} \Big[ g^{\sigma\nu} (p_1\cdot p_2)
          - p_1^{\sigma}\,p_2^{\nu} \Big] + g^{\nu\rho}
          \Big[ g^{\sigma\mu} (p_1\cdot p_2) - p_1^{\sigma}\,p_2^{\mu} \Big] \Big)
\nonumber\\
&\!\!+\!\!&
B_V\cdot \Big( g^{\sigma\rho} \Big[ p_1^{\mu}\,p_2^{\nu} + p_1^{\nu}\,p_2^{\mu} \Big]
         - g^{\mu\sigma} p_1^{\nu}\,p_2^{\rho} - g^{\nu\sigma} p_1^{\mu}\,p_2^{\rho} \Big)
         \,,
\label{eq:amp3}
\ena
where the two independent form factors
$A_V(g_{X},g_{J/\psi},g_{V},p^2,p_1^2,p_2^2)$ and
$B_V(g_{X},g_{J/\psi},g_{V},p^2,p_1^2,p_2^2)$ are determined according
to Eq.~(\ref{eq:amplitude}).

Consequently, we calculate
\be
| {\cal M}_{X_2 JV} |^2 \sim
| \varepsilon_{\mu\nu}(p)  \, \varepsilon^{\ast}_{\rho}(p_1) \,
\varepsilon^{\ast}_{\nu}(p_2) \, \, T_{X_2 JV}^{\mu\nu\rho\sigma}|^2
\!=\!  M_{X_2}^4
\big( C^V_{A} \!\cdot\! A_V^2 \!+\! C^V_{AB}\!\cdot\! A_V \!\cdot\! B_V
\!+\! C^V_{B}\!\cdot\! B_V^2 \big) \,,
\label{eq:matrixX2}
\en
where the coefficients $C^V_{A}$, $C^V_{AB}$ and $C^V_{B}$ are completely
defined through the meson masses as follows:
\bea
C^V_{A}
&\doteq&
 [ (3 + 2 \xi_V) \, \xi_J^4+ (1 - \xi_V)^4  \, (3 +
      2 \xi_V) + (28 + 20 \, \xi_V - 8 \, \xi_V^2) \, \xi_J^3            \nonumber\\
& + &
   4 \, \xi_J (1 - \xi_V)^2 (7 + 5 \, \xi_V - 2 \xi_V^2)
   - 2 \, \xi_J^2 (31 - 52 \, \xi_V + 27 \, \xi_V^2 - 6 \, \xi_V^3) ] / (12 \, \xi_J) \,,
\nonumber\\
C^V_{AB}
&\doteq&
[\xi_J^4 (7 - 2 \,\xi_V) - (1 - \xi_V)^4 (3 + 2 \, \xi_V)
 - 2 \xi_J^3 (9 + 5 \, \xi_V - 4 \, \xi_V^2)  \nonumber\\
& + &
   2 \, \xi_J (1 - \xi_V)^2 (1 + 5 \, \xi_V + 4 \, \xi_V^2)
 + 4 \, \xi_J^2 (3 - \xi_V + \xi_V^2 - 3 \, \xi_V^3)]  / (6 \, \xi_J) \,,
\nonumber\\
C^V_{B}
&\doteq&
[3 + 4 \, \xi_J + 2 \, \xi_V) (\xi_J^2 + (1 - \xi_V)^2 -
   2 \, \xi_J (1 + \xi_V)]^2  / (12 \, \xi_J) \,,
\label{eq:CAACABCBB}
\ena
where $\xi_J \doteq M^2_{\Jpsi}/M^2_{X_2}$ and
$\xi_V \doteq M^2_{V}/M^2_{X_2}$ for $V=\{ \rho^0, \omega \}$.

The two-body strong decay width reads:
\be
\Gamma_{X_2 JV} = \frac{1}{2S+1} \, \frac{|\vec{p}_2|}{8\pi M^2_{X_2}} \,
\sum_{polar} | {\cal M}_{X_2 JV} |^2  \,,
\label{eq:widXJV}
\en
where
$|\vec{p}_2|\doteq\lambda^{1/2}(M^2_{X_2},M^2_{\Jpsi},M^2_{V})/(2M_{X_2})$
and $\lambda(x,y,z)\doteq x^2+y^2+z^2-2\,(xy+xz+yz)$ is the K\"all\'en
kinematical function while $S=2$ is the spin value of $X_2$.

\section{Numerical Results}
\label{sec:numerical}

Many experimental and theoretical investigations are devoted to understand
the nature of the exotic $XYZ$ family and considerable results have been
achieved to determine the  mass, width, and main quantum numbers (for reviews
see, e.g. \cite{brambilla:2020pr,lebed:2017ppn,chen:2016pr}).

In particular, a recent observation of  a structure in the invariant mass
distribution of the $\gamma \psi(2S)$ with a mass of $4014.3\pm 4.0 \pm 1.5$
~MeV and a width of $4\pm 11\pm 6$ MeV  by the Belle collaboration
\cite{wang:2022belle} has became a stimulating motive
in the investigation of the exotic $XYZ$ states despite a low global
significance of $2.8\sigma$ of the new structure.

In this Section, we systematically continue our investigation of  the hidden
charmonium decays of the exotic $XYZ$ states and consider the two-body
strong decays of the spin-2 partner of $X(3872)$, namely $X_2 \to \rho^0 \, J/\psi$
and  $X_2 \to \omega \, J/\psi$. We assume a four-quark content of $X_2$ with
a pure $D^*\bar{D}^*$  mesonic molecule structure. In doing so, we consider
only the LO contributions to the decays corresponding to the two-leaf Feynman
diagram shown in Fig.~\ref{fig:FIG2} in the framework of the CCQM approach.

The model parameters in the CCQM are determined by minimizing $\chi^2$ in fits
to the latest available experimental data and some lattice results. The fitted
parameters may vary around their central value by about few percents, and the
errors of our calculations do not exceed $\pm 10 \%$ percents.

The most of hadron 'size' parameters in the CCQM are more or less rigidly fixed
by fitting to available experimental data (see, e.g.
in~\cite{Dubnicka:2020xoh,Ganbold:2021nvj}). The updated central values of the
basic CCQM parameters, namely, the universal infrared cutoff parameter
$\lambda$, the constituent quark masses ($m_{u/d}, m_c$) and the hadron 'size'
parameters ($\Lambda_H$)  are partly shown in Table~\ref{tab:TAB3}.
We proceed with our computations using these values, which remain unaltered.

\begin{table}[h]
\caption{Model basic parameter values (in GeV).}
\label{tab:TAB3}
\begin{tabular}{ c  c  c  c  c  c }
\hline
$\quad \lambda$ & $\quad m_{u/d}$ & $\quad m_c$ & $\quad \Lambda_{\rho^0}$
& $\quad \Lambda_{\omega}$   & $\quad \Lambda_{\Jpsi}$ \\
\hline
\quad 0.181  & \quad 0.241    & \quad  1.670  & \quad  0.61 & \quad  0.80  & \quad  1.55 \\
\hline
\end{tabular}
\end{table}

\begin{figure}[H]
\begin{center}
\includegraphics[scale=0.35]{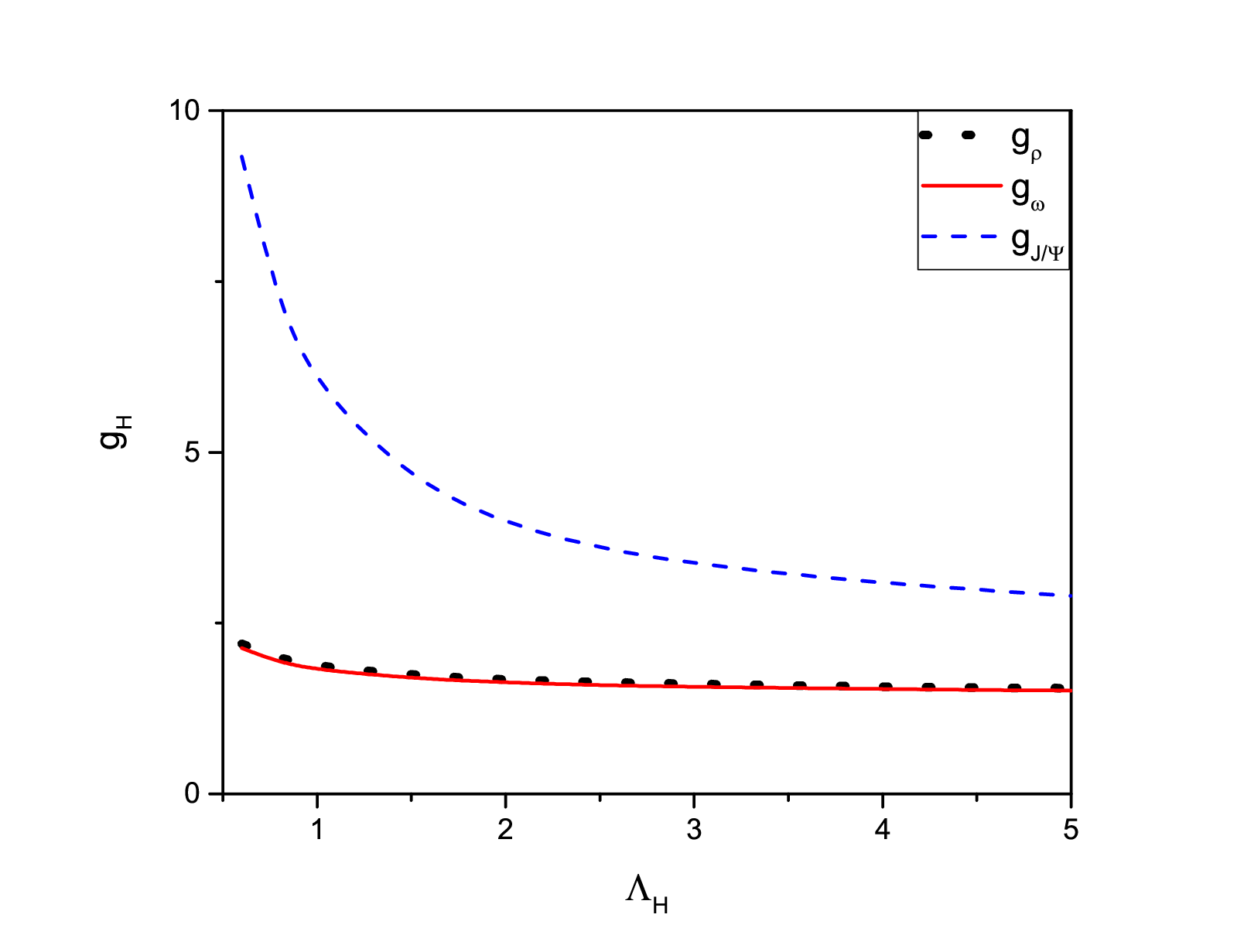}
\end{center}
\caption{
The renormalized couplings $g_H$ of the mesons ($H=\{ \rho^0,\omega,\Jpsi \}$)
calculated in dependence on the 'size' parameters $\Lambda_H$ (in GeV).
}
\label{fig:FIG3}
\end{figure}

First, we calculate the renormalized couplings $g_{H}$ of the participating
mesons ($ H=\{\Jpsi, \omega, \rho^0 \}$) according to Eq.~(\ref{eq:renorm}).
The numerical values of  $g_{H}$ estimated in dependence on the 'size'
parameters $\Lambda_{H}$ are given in Fig.~\ref{fig:FIG3}.

We change the only totally adjustable size parameter $\Lambda_{X_2}$ at the
appropriate intervals for the numerical evaluation of the strong decay widths
of $X_2$.

Remember, we utilized the 'size' parameter value of the tetraquark state
X(3872) in the interval $\Lambda_X\in [3.0 \div 4.0]$ GeV in our previous study
\cite{dubnicka2011}.  More recently, the 'size' value for a heavier exotic state
$Y(4230$ has been corrected to $\Lambda_Y\in [5.15 \pm 0.26]$ GeV in
\cite{Ganbold:2024epja}.

Because the physical mass of the $X_2$ state is between two masses,
$M_X=3.872$ GeV and $M_Y=4.222$ GeV, we will search for the necessary
parameter $\Lambda_{X_2}$ in the range of (3.5 - 5.5)~GeV by adhering to
the model's pattern - 'the heavier a hadron, the larger its size'.

Then, using Eq.~(\ref{eq:widXJV}), we can estimate the hidden charm partial
decay widths of the two-body strong decays of $X_2$, once the renormalization
couplings $g_H$ have been computed.

{\bf 1}.  Let us first make a simple and rough approximation to the desired
branching ratio $\mathrm{BR_{X_2}}$. We easily calculate the ratios between
the relevant phase spaces and corresponding renormalized couplings as follows:
\be
\frac{ \lambda^{1/2}( M^2_{X_2}, M^2_{\Jpsi}, M^2_{\omega} ) }
       { \lambda^{1/2}( M^2_{X_2}, M^2_{\Jpsi}, M^2_{\rho^0} ) }
\approx 0.976\,,
\qquad
\frac{g_{\omega}^2}{g_{\rho^0}^2} \approx 0.780\,.
\label{eq:rate1}
\en
Then, by neglecting the difference between the corresponding form factors
$C_*^\omega$ and $C_*^{\rho^0}$ (* = \{A,AB,B\}) defined in
Eq.~(\ref{eq:CAACABCBB}), one can approximately write down the ratio:
\begin{equation}
\mathrm{BR_{X_2}^{approx}}  \approx 0.762
\label{eq:BWapprox}
\end{equation}
that is less than unity.

\begin{figure}[H]
\begin{center}
\includegraphics[scale=0.4]{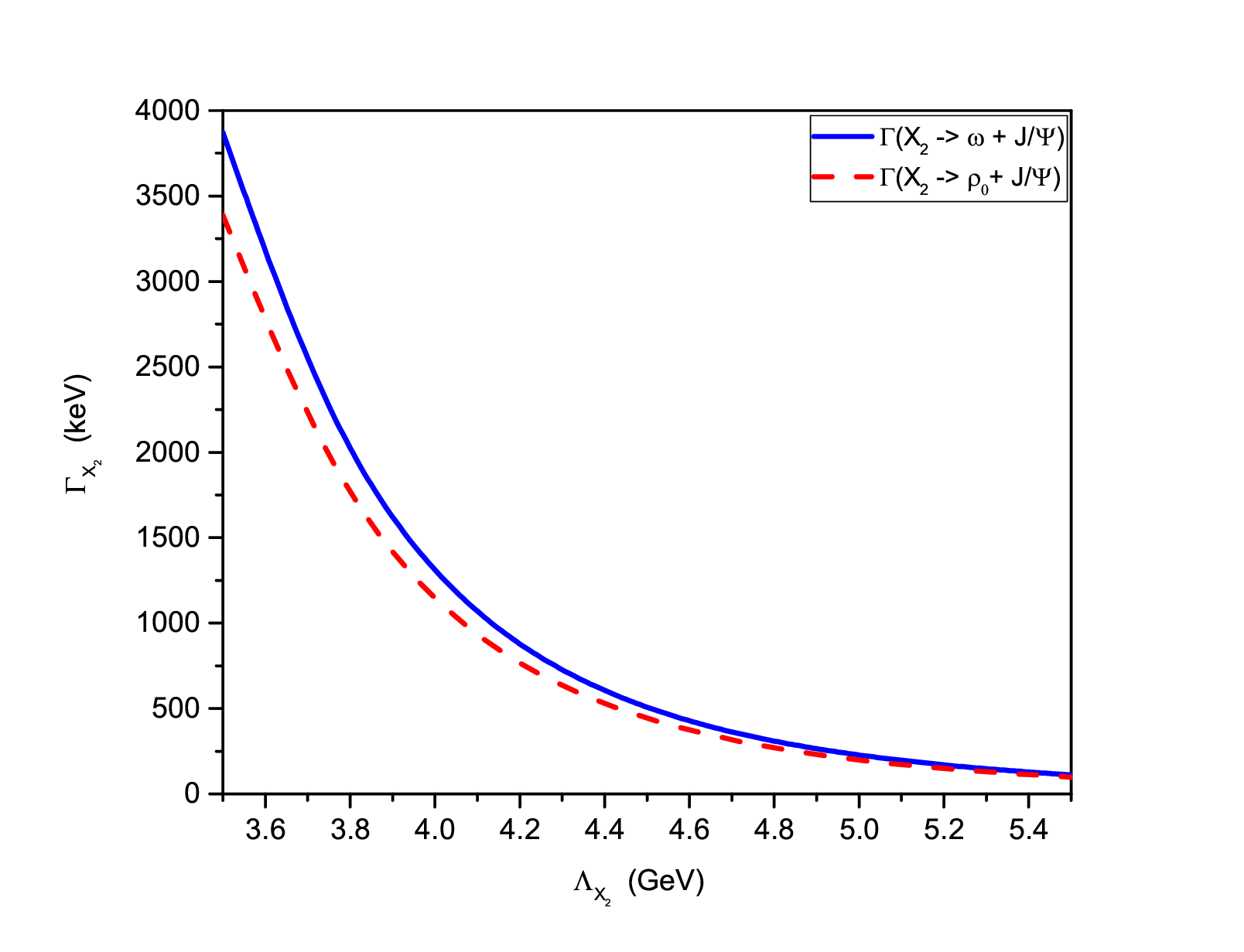}
\end{center}
\caption{
The dependencies of the partial decay widths $\Gamma(X_2 \to \omega \, J/\psi)$
and $\Gamma(X_2 \to \rho^0 \, J/\psi)$ on the 'size' parameter $\Lambda_{X_2}$
for a fixed central value of the exotic state mass $M_{X_2} = 4.014$ GeV.
}
\label{fig:FIG4}
\end{figure}

{\bf 2}. Now we take into account accurately the real contributions of the
matrix elements and calculate the partial decay widths defined in
Eq.~(\ref{eq:widXJV})  in dependence on the 'size' parameter $\Lambda_{X_2}$.
The obtained results are represented in Fig.~\ref{fig:FIG4} for a fixed central
value of the exotic state mass $M_{X_2} = 4.014$ GeV.

One can see in Fig.~\ref{fig:FIG4} that both the partial decay widths decrease
monotonically as the 'size' parameter $\Lambda_{X_2}$ increases.

i) Accordingly, if we admit that the partial two-body strong decay widths
$\Gamma(X_2 \to \omega \, J/\psi)$ and $\Gamma(X_2 \to \rho^0 \, J/\psi)$ are
of hundreds $keV$, then our estimates impose a certain restriction on the
allowed 'size' parameter within an interval $\Lambda_{X_2} \in [4.0 \div 5.5 ]$ GeV.

ii) However, the corresponding ratio of these partial widths
\be
\mathrm{BR_{X_2}^{CCQM}}    \doteq
\frac{\Gamma(X_2 \to \omega  J/\psi)} {\Gamma(X_2 \to \rho^0  J/\psi)}
= 1.143 \sim 1.147 \qquad \mathrm{for}
\qquad \Lambda_{X_2} \in [3.5 \,, \, 5.5 ] \,\, \mathrm{GeV}
\label{eq:BRstr2}
\en
almost cancels or at least weakens the 'size'-dependence.

{\bf 3}. Recently, by assuming the $X_2$ as a pure molecule of the
$D^*\bar{D}^*$, the hidden charmonium decays of the $X_2 \to \omega J/\psi $
and $X_2 \to \rho^0 J/\psi$ via the intermediate meson loops have been
estimated in a framework of the effective field theory \cite{zheng:2024prd}.

\begin{figure}[h]
\begin{center}
\includegraphics[scale=0.4]{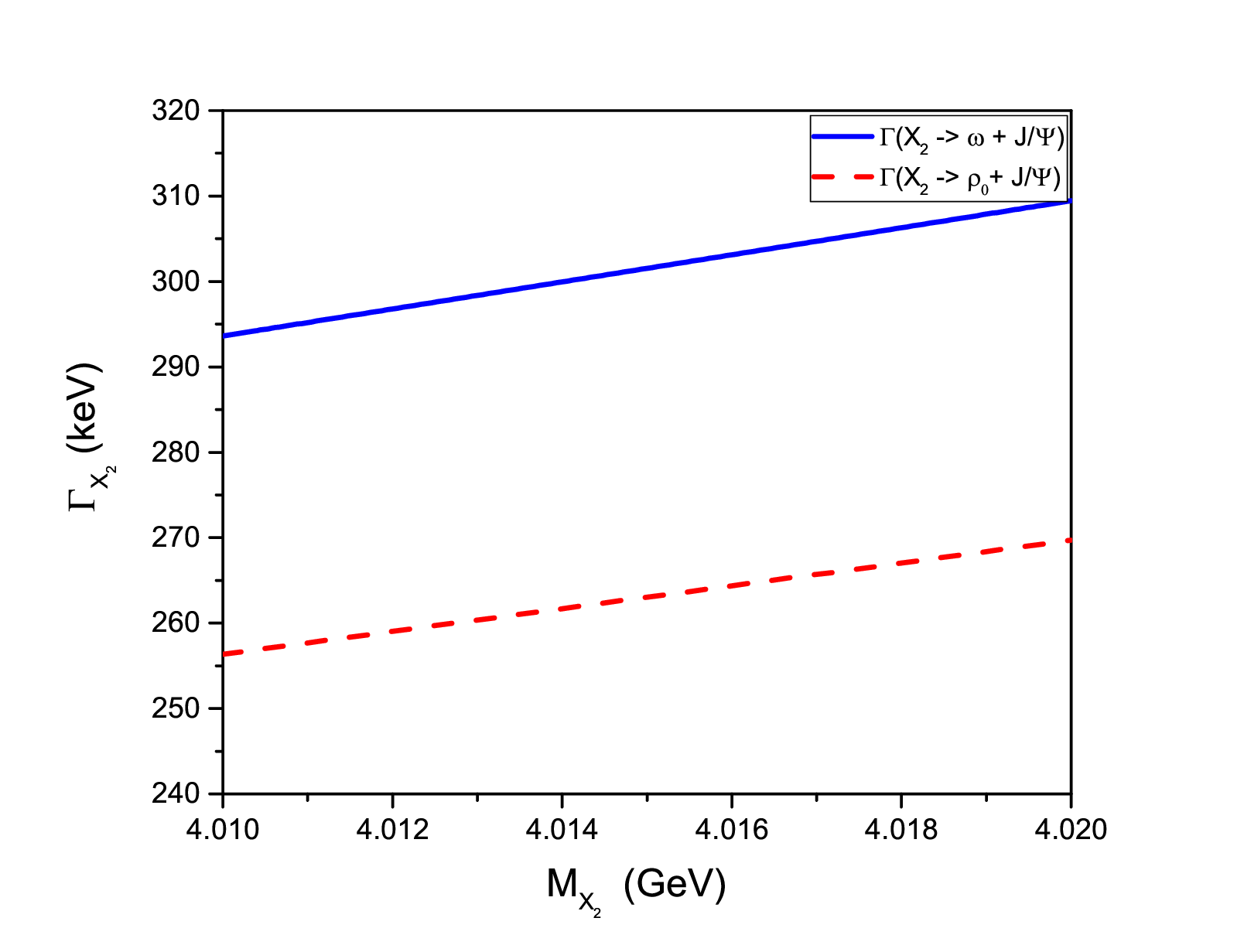}
\end{center}
\caption{
The dependencies of the partial decay widths $\Gamma(X_2 \to \omega \, J/\psi)$
and $\Gamma(X_2 \to \rho^0 \, J/\psi)$ on the exotic state mass $M_{X_2}$
for a fixed 'size' parameter value $\Lambda_{X_2} = 4.80$ GeV.
}
\label{fig:FIG5}
\end{figure}

The results indicated that the decay widths are strongly dependent on the $X_2$
mass. In particular, within a short interval for the mass of $X_2$ ranging from
$4.009$ GeV to $4.020$ GeV,  the width ranges from $\sim 0.2$ to $200
~\mathrm{keV}$ for $X_2 \to \rho^0 J/\psi$, while for $X_2 \to \omega J/\psi$
it is between 0.1 and 1.5 MeV. Hereby,  the width for $X_2 \to \rho^0 J/\psi$
first increases with the $X_2$ mass to a peak value at $m_{X_2} = 4.0137
~\mathrm{GeV}$, then drops until $\sim 4.0167~\mathrm{GeV}$, and finally starts
to increase again. However, the width for $X_2 \to \omega J/\psi $ shows
opposite variations with the $X_2$ mass at a given $\alpha$, i.e., the width
exhibits a valley at $m_{X_2}=4.0137$~GeV, while near $m_{X_2}=4.0175$~GeV
it is a peak. At present center value of the  mass $4014.3$~MeV, the width for
the $X_2 \to \rho^0 J/\psi$ is predicted to be a few tens of keV, while it is
on the order of $10^{2\text{-}3}$~keV for the $X_2 \to \omega \Jpsi$
\cite{zheng:2024prd}.

We have also investigated the dependence of the partial decay widths
$\Gamma(X_2 \to \omega J/\psi)$ and $\Gamma(X_2 \to \rho^0 J/\psi)$ on the
exotic hadron mass deviation within the experimental uncertainty reported in
\cite{Belle:2021nuv}. The corresponding numerical results are shown in
Fig.~\ref{fig:FIG5} for a given value of 'size' parameter
$\Lambda_{X_2}=4.80$~GeV.

Our numerical results indicate that the decay widths slowly increase without
any peaks and drops. In the mass interval from $4.010$ GeV to $4.020$ GeV,
the decay width ranges from $294$ to $309$ keV for $X_2\to \omega \, J/\psi$,
while for $X_2\to \rho^0\, J/\psi $ it is between $257$ and $270$ keV, by keeping
the branching ratio constant $\mathrm{BR_{X_2}^{CCQM}} = 1.14$.

This result is not surprisingly. The hidden charm decays of the $X_2$ in our
approach occur via the confined quark loops but not the charmed $D^*$ and
$\bar{D}^*$ meson loops, without any threshold effects.

\section{Summary}
\label{sec:summary}

The hidden-charm strong decays of the spin-2 exotic charmonium-like state
$X_2(4014)$ into channels $\omega J/\Psi$ and $\rho^0 J/\Psi$ have been
studied within the framework of the covariant confined quark model, designed 
to eliminate any UV divergences in quark loops.

We have interpreted the exotic hadron $X_2$ as a four-quark state with a 
$D^*\bar{D}^*$ molecular-type interpolating quark current and computed the 
leading-order strong decay widths at the level of two-petal quark-loop diagrams.

We have calculated the partial widths of the above-mentioned decay modes and
their branching ratio recently discussed in literature~\cite{shi:2023,zheng:2024prd}.
In our calculation, we used the average mass and full decay width of the $X_2$ 
state reported by the Belle Collaboration~\cite{Belle:2021nuv}.

\vspace*{2mm}

Our findings are:

{\bf i}. 
The partial strong-decay widths $\Gamma(X_2 \to \omega \, J/\psi)$ and
$\Gamma(X_2 \to \rho^0 \, J/\psi)$ depend significantly on the model size 
parameter $\Lambda_{X_2}$, decreasing rapidly from 3 - 4 MeV to several 
hundreds of keV in the reasonable interval (3.5 - 5.5 GeV) for a fixed mass 
$M_{X_2}=4.014$ GeV. 

{\bf ii}.  
We may calculate a simple and rough approximation of the corresponding 
branching ratio by taking into account only effects due to the phase space 
factors and renormalized couplings as follows:
$\mathrm{ BR_{X_2}^{approx}} \approx 0.762$. 

However, our accurate numerical results represented in Fig.~\ref{fig:FIG5}
clearly show that the ratio $\mathrm{ BR_{X_2}^{CCQM}} \simeq 1.14$ is larger 
than unity and almost independent of $\Lambda_{X_2}$. This may indicate a 
relative dominance of the $\omega \, J/\psi$ decay mode.

{\bf iii}. 
We have also investigated the sensibility of our numerical results on the 
exotic hadron mass deviation under the data uncertainty reported 
in~\cite{Belle:2021nuv}. Remember that the corresponding numerical results 
for the decay widths and the ratio reported in~\cite{shi:2023,zheng:2024prd} 
demonstrated very strong dependencies on the $X_2$ mass value.

Our numerical results show that the decay widths monotonically increase
with no peaks and drops in the mass interval from $4.010$ GeV to 
$4.020$ GeV;  the width ranges from $294$ to $309$ keV for 
$X_2\to J/\psi \omega$, while for $X_2\to J/\psi \rho^0$, it is between $257$
and $270$ keV, by keeping the branching ratio almost constant
($\mathrm{BR_{X_2}^{CCQM}} = 1.14$).

The reason is simple: the hidden charm decays of the $X_2$ in our approach
occur via the confined quark loops but not the charmed $D^*$ and $\bar{D}^*$
meson loops, so with no threshold effects.

\vspace*{2mm}

By conclusion, the exotic spin-2 state $X_2$ has been interpreted as a
four-quark state of molecular-type interpolating current within the CCQM.
We have calculated the strong decay (to $\omega J/\psi$ and $\rho^0 J/\psi$)
widths of the exotic state and estimated the corresponding branching ratio.

We have maintained the basic CCQM parameters in our calculation,  adding 
only one new tunable size parameter ($\Lambda_{X_2}$) to represent the quark 
distribution inside the hadron.

Our numerical results show that the strong decay widths under consideration
change smoothly, but their ratio $\mathrm{BR_{X_2}^{CCQM}} = 1.14$ is almost
constant in the wide and reasonable interval .

Our numerical results indicate that the strong decay widths under consideration 
vary smoothly, but within the wide and tolerable interval 
$\Lambda_{X_2}\in [3.5-5.5]$~GeV, their ratio is nearly constant 
($\mathrm{BR_{X_2}^{CCQM}} = 1.14$).

Additionally, it was discovered that the decay widths under investigation 
showed only a minor sensitivity to the mass variation of $X_2$ determined in 
the most recent experiment~\cite{Belle:2021nuv}.

We draw the conclusion that the estimated branching ratio and hidden-charm 
strong-decay widths are in good accord with the most recent theoretical 
predictions in~\cite{shi:2023,zheng:2024prd} and may support the four-quark 
molecular-type structure of $X_2$.

We anticipate that further, more accurate experimental data on the $X_2$'s 
particle characteristics will enable us to refine the model's parameters and 
make more intelligible deductions regarding the underlying workings of this 
exotic state.



\end{document}